\documentclass[11pt,a4paper]{article}
\usepackage{jheppub}
\usepackage{amsmath}
\usepackage{amsfonts}
\usepackage{amssymb}
\usepackage{physics}
\usepackage{graphicx}
\graphicspath{{figures/}}
\usepackage{caption}
\usepackage{subcaption}

\usepackage{tikz}
\usepackage{feynmp-auto}
\usetikzlibrary{calc}

\usepackage[T1]{fontenc} % \textsc{} in \section{}

\title{\boldmath Enhancing the Sensitivity for Triple Higgs Boson Searches with Deep Learning Techniques}

\author[a,b]{Cheng-Wei Chiang,}
\author[a]{Feng-Yang Hsieh,}
\author[c]{Shih-Chieh Hsu,}
\author[d,e]{Ian Low}
\author[e]{and Zhi-Zhong Li}

\affiliation[a]{\vspace{0.2cm}Department of Physics, National Taiwan University, Taipei 10617, Taiwan}
\affiliation[b]{Physics Division, National Center for Theoretical Sciences, Taipei 10617, Taiwan}
\affiliation[c]{Department of Physics, University of Washington, Seattle, WA 98195, U.S.A.}
\affiliation[d]{High Energy Physics Division, Argonne National Laboratory, Lemont, IL 60439, U.S.A.}
\affiliation[e]{Department of Physics and Astronomy, Northwestern University, Evanston, IL 60208, U.S.A.\\}

\emailAdd{chengwei@phys.ntu.edu.tw}
\emailAdd{f10222035@ntu.edu.tw}
\emailAdd{schsu@uw.edu}
\emailAdd{ilow@northwestern.edu}
\emailAdd{Zhi-zhongLi2024@u.northwestern.edu}

\abstract{Using two benchmark models containing extended scalar sectors beyond the Standard Model, we study deep learning techniques to enhance the sensitivity of resonant triple Higgs boson searches in the fully hadronic $6b$ channel, which suffers from the combinatorial challenge of reconstructing the Higgs bosons correctly from the multiple $b$-jets.  More specifically, we employ the framework of Symmetry Preserving Attention Network (\textsc{Spa-Net}), which takes into account the permutational symmetry when a correct pairing of $b$-jets is achieved, to tackle both jet pairing and event classification. Significantly improved efficiency is achieved in signal and background discrimination. When comparing with the conventional Dense Neural Networks, \textsc{Spa-Net} results in up to 40\% more stringent limits on resonant production cross-sections. These results highlight the potential of using advanced machine learning techniques to significantly improve the sensitivity of triple Higgs boson searches in the fully hadronic channel.}

\begin{document}

\maketitle

\section{Introduction}% (fold)
\label{sec:introduction}

    Since the discovery of the 125~GeV Higgs boson, an important task is to determine the Higgs potential, particularly its self-interactions. It is well known that the Higgs potential in the Standard Model (SM) is insufficient to induce a strong first-order electroweak phase transition, which is a necessary condition for electroweak baryogenesis and the generation of the observed matter-antimatter asymmetry~\cite{Sakharov:1967dj, Kuzmin:1985mm, Cohen:1993nk}. 

    New physics may modify the Higgs potential at finite temperatures and enable a strong first-order phase transition. This often leads to deviations in the Higgs self-couplings from their SM values. In particular, the trilinear coupling affects the di-Higgs production cross-section and kinematics at the LHC~\cite{Kanemura:2004ch}. Measurements of di-Higgs production thus provide indirect access to the Higgs potential~\cite{Dawson:2015oha, Chen:2014xra}. Both ATLAS~\cite{ATLAS:2024ish} and CMS~\cite{CMS:2024awa} have placed constraints on the trilinear coupling using various Higgs decay channels.

    Triple Higgs production is known to occur at a very low rate but provides complementary information on the trilinear coupling, which could be used in combination with the above searches. Furthermore, it can provide the first experimental constraint on the quartic Higgs coupling. However, due to its extremely small cross-section, obtaining a meaningful measurement requires both high luminosity and improved analysis techniques.

    In this study, we investigate the resonant production of three SM-like Higgs bosons in models such as the Dark Matter and CP Violation (DM-CPV) singlet model~\cite{Chen:2022vac} and the Two Real Singlet Model (TRSM)~\cite{Robens:2019kga}. Each Higgs boson decays into a pair of $b\overline{b}$, resulting in the $6b$ final state. Given the Higgs boson mass of $\text{125 GeV}$, this decay channel has the highest branching ratio among all possible modes. Nevertheless, the $6b$ final state faces significant challenges due to the overwhelming QCD multi-jet background. Furthermore, because of the inability to distinguish a $b$-jet from a $\bar b$-jet, it is a very challenging task experimentally to form the correct pairing among the final $b$-jets to reconstruct the Higgs mass and the associated kinematic distributions accurately. 

	To perform the jet pairing task and construct the Higgs boson candidate, conventional methods generally adopt cut-based approaches. These methods enumerate all possible pairing configurations and calculate a specific metric for each. The optimal configuration is then selected by minimizing or maximizing this quantity. While these methods are straightforward to implement, they suffer from significant computational expense due to the combinatorial explosion in possible pairings and fail to incorporate the intrinsic label symmetries present in jet assignment tasks.

    Following jet pairing, event classification is typically performed using either sequential kinematic cuts or machine learning techniques. Previous work has shown that even basic dense neural networks (Dense-NNs) can significantly enhance signal sensitivity in di-Higgs analyses~\cite{Amacker:2020bmn}, and it has been adopted in recent ATLAS tri-Higgs searches~\cite{ATLAS:2024xcs}. In addition, CMS has recently also pursued tri-Higgs searches in the $4b2\gamma$ channel~\cite{CMS:2025jkb}.

	Motivated by these considerations, we propose to employ machine learning algorithms based on deep neural networks to enhance the sensitivity of experimental tri-Higgs searches in the $6b$ channel. Specifically, we explore the use of a novel neural network architecture known as the Symmetry Preserving Attention Network (\textsc{Spa-Net})~\cite{PhysRevD.105.112008, 10.21468/SciPostPhys.12.5.178, Fenton:2023ikr}, which is capable of simultaneously performing signal / background separation and identifying the correct pairings among the $b$-jets in the final states. 

    \textsc{Spa-Net} is specifically designed to incorporate symmetry considerations inherent in jet assignment tasks. For the di-Higgs analysis, it has been demonstrated that \textsc{Spa-Net} outperforms~\cite{Chiang:2024pho} existing experimental techniques employed in the $4b$ channel \cite{ATLAS:2018rnh, ATLAS:2022hwc, ATLAS:2023qzf, CMS:2018qmt, CMS:2022cpr}, as well as Dense-NN analyses \cite{Amacker:2020bmn}, in terms of sensitivity improvement. For the tri-Higgs case, recent studies have also shown that \textsc{Spa-Net} can improve the jet pairing performance~\cite{Li:2024qfq}. In this work, we extend its application to the fully resolved $6b$ final states, aiming to establish its effectiveness in this regime and evaluate its potential to improve both classification and cross-section sensitivity.

    This paper is organized as follows. In section~\ref{sec:sample_preparation}, we describe two models considered in this work, following by a description of the signal and background sample preparation and simulation settings. Section~\ref{sec:jet_pairing} details the jet pairing methods, emphasizing \textsc{Spa-Net}'s advantages. In section~\ref{sec:neural_network_classifier}, we compare the \textsc{Spa-Net} and Dense-NN classifiers, highlighting the benefits of \textsc{Spa-Net} architecture. Section~\ref{sec:results} showcases the training outcomes and the resulting cross-section upper limits, demonstrating the superior performance of \textsc{Spa-Net}.  Finally, we conclude with a discussion of future prospects.
 
% section introduction (end)

\section{Framework and sample preparation} % (fold)
\label{sec:sample_preparation}

    For LHC experiments running at $\sqrt{s}=14~\mathrm{TeV}$, the Standard Model prediction for tri-Higgs production is extremely small, with a cross-section of about $0.103~\mathrm{fb}$ at NNLO~\cite{deFlorian:2019app}.  Models with an extended Higgs sector often proffer resonant contributions to the tri-Higgs production, possibly enhancing its cross-section by a few orders of magnitude.  It thus serves as a good channel to probe new physics in the Higgs sector.

    A simple and well-motivated example is a singlet model with dark matter and CP violation (DM-CPV)~\cite{Chen:2022vac}, although it is possible to generate the triple Higgs boson signature in a CP-violating two-Higgs-doublet-model as well~\cite{Low:2020iua}. In this model, the scalar sector is extended with a complex singlet $S$, which, after mixing with the physical neutral component of the SM doublet $H$, leads to three neutral Higgs mass eigenstates $h_1, h_2,$ and $h_3$, with $h_1$ identified as the observed 125~GeV Higgs boson. At the Lagrangian level, the renormalizable scalar potential before electroweak symmetry breaking is
    \begin{align}
        V(H,S) =&~ \mu^{2}H^{\dagger}H + \lambda(H^{\dagger}H)^{2} + \frac{\delta_{1}}{4}H^{\dagger}HS + \frac{\delta_{2}}{2}H^{\dagger}H|S|^{2} + \frac{\delta_{3}}{4}H^{\dagger}HS^{2} \nonumber\\  
        & + \frac{b_{1}}{4} S^{2} + \frac{b_{2}}{2} |S|^{2} + \frac{c_{1}}{6} S^{3} + \frac{c_{2}}{6} S|S|^{2} + \frac{d_{1}}{8} S^{4} + \frac{d_{2}}{4} |S|^{4} + \frac{d_{3}}{8} S^{2}|S|^{2} + \text{h.c.}
    \end{align}
    Expanding around the vacuum expectation values (VEVs) of the singlet and the doublet and rotating to the mass basis yield the interactions of cubic or higher powers in the form
    \begin{equation}
        V \supset \frac{1}{3!}\sum_{i,j,k=1}^3 g_{ijk}\,h_i h_j h_k + \text{(higher-power terms)} ,
        \label{eq:cubic terms}
    \end{equation}
    which can mediate resonant cascade decays and give rise to the following process:
    \begin{equation}
        gg \;\to\; h_3 \;\to\; h_2 h_1 \;\to\; h_1 h_1 h_1 .
    \end{equation}

    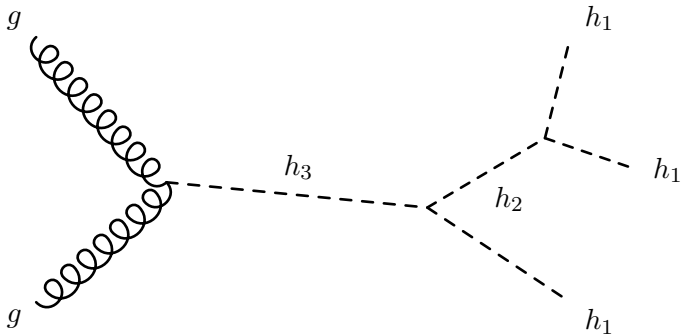
\begin{figure}[htb]
        \centering
        \begin{fmffile}{gg_to_3higgs}
        \begin{fmfgraph*}(250,100)
            \fmfleft{i1,i2}
            \fmfright{o1,o2,o3}
        
            % Vertices
            \fmf{gluon}{i1,v1}
            \fmf{gluon}{i2,v1}
        
            \fmf{dashes,label=$h_3$}{v1,v2}
            \fmf{dashes,label=$h_2$}{v2,v3}
            \fmf{dashes}{v2,o1}
            \fmf{dashes}{v3,o2}
            \fmf{dashes}{v3,o3}
        
            % Labels
            \fmflabel{$g$}{i1}
            \fmflabel{$g$}{i2}
            \fmflabel{$h_1$}{o1}
            \fmflabel{$h_1$}{o2}
            \fmflabel{$h_1$}{o3}
        \end{fmfgraph*}
        \end{fmffile}
        \vspace{20pt}
        \caption{Feynman diagram for resonant triple Higgs production via gluon fusion in the TRSM, where the heavy scalar $h_3$ decays through $h_2$ to produce three SM-like Higgs bosons.}
        \label{fig:gghhh}
    \end{figure}

    A similar mechanism also appears in another widely considered model, the Two Real Singlet Model (TRSM)~\cite{Robens:2019kga, Papaefstathiou:2020lyp}. At the Lagrangian level, the scalar sector is extended with two real singlets $S$ and $X$, with each subject to a separate $\mathbb{Z}_2$ symmetry, leading to the potential
    \begin{align}
        V =&~ \mu^2 H^\dagger H + \lambda (H^\dagger H)^2 + \mu_S^2S^2 + \lambda_SS^4 + \mu_X^2X^2 + \lambda_XX^4 \nonumber\\
        & + \lambda_{H S}H^\dagger HS^2 + \lambda_{H X}H^\dagger HX^2 + \lambda_{SX}S^2 X^2 ,
    \end{align}
    where all the parameters are real. After electroweak symmetry breaking and rotation to the mass basis, this potential also lead to interactions of the same form as in Eq.~\eqref{eq:cubic terms}.  One important difference between the DM-CPV model and the TRSM is that the couplings $g_{ijk}$ are generally complex in the former case and purely real in the latter one. Ignoring CP violation in the current analysis, both models would give rise to the same event kinematics~\cite{ATLAS:2024xcs}. In the following, we will use the TRSM to generate events, for the ease of a more direct comparison with the published ATLAS results. In our simulation, the SM-like Higgs $h_1$ decays predominantly into $b\bar{b}$, resulting in an all-hadronic $6b$ final state. 
    
    We select five benchmark points from Ref.~\cite{ATLAS:2024xcs} for our analysis. The corresponding masses and leading-order (LO) cross-sections are summarized in table~\ref{tab:trsm_xsec}. Calculations at NNLO+NNLL suggest that gluon-gluon fusion production of $h_3$ receives a $k$-factor of $\simeq 2.5$~\cite{Papaefstathiou:2020lyp}.

    \begin{table}[htpb]
        \centering
        \caption{Cross-sections of $gg \to h_3 \to h_2 h_1 \to h_1 h_1 h_1$ at $\sqrt{s} = 13$~TeV and 14~TeV for five benchmark points of the TRSM. The cross-sections are computed using \texttt{MadGraph}~\cite{Alwall:2014hca} at the leading order. The mass of $h_1$ is fixed to 125~GeV for all benchmarks.}
        \label{tab:trsm_xsec}
        \begin{tabular}{c|cc|cc}
            \hline \hline
                                       & \multicolumn{2}{c|}{TRSM} & \multicolumn{2}{c}{DM-CPV} \\
            $(m_{h_3}, m_{h_2})$ [GeV] & 13~TeV [fb] & 14~TeV [fb] & 13~TeV [fb] & 14~TeV [fb] \\
            \hline
            (420, 280)                 & 47.26       & 55.79       & 9.32        & 11.10       \\
            (500, 275)                 & 37.34       & 44.51       & 4.52        & 5.55        \\
            (500, 300)                 & 36.66       & 43.71       & 4.38        & 5.20        \\
            (520, 325)                 & 30.44       & 36.39       & 3.16        & 3.77        \\
            (500, 350)                 & 27.53       & 32.85       & 3.26        & 3.89        \\
            \hline \hline
        \end{tabular}
    \end{table}

    The main background of the $6b$ signature arises from QCD multi-jet production. In this work, we only consider the dominant process $pp\to b\overline{b}b\overline{b}b\overline{b}$. Subleading contributions come from processes in which light-flavor or charm jets are misidentified as $b$-jets. Based on our simulations, these are found to contribute less than 10\% of the dominant process and are therefore neglected in this analysis.
    
    We utilize \verb|MadGraph 3.3.1|~\cite{Alwall:2014hca} to generate signal and background samples at a center-of-mass energy of $\sqrt{s} = 13$ TeV with the \verb|NNPDF23_nlo_as_0119| PDF set~\cite{Ball:2012cx}. All produced Higgs bosons forced to decay into $b\bar{b}$ pairs.  Parton showering and hadronization are simulated using \verb|Pythia 8.306|~\cite{Sjostrand:2014zea} with \verb|NNPDF2.3 LO| PDF set. We use \verb|Delphes 3.4.2|~\cite{deFavereau:2013fsa} for fast detector simulations. Jet reconstruction is performed with \verb|FastJet 3.3.2|~\cite{Cacciari:2011ma} using the anti-$k_t$ algorithm~\cite{Cacciari:2008gp} and a jet radius of $R = 0.4$. These jets are required to have
    \begin{enumerate}
        \item transverse momenta $p_{\text{T}} > 20$~GeV, and
        \item pseudorapidities $|\eta| < 2.5$.
    \end{enumerate}
    
    To ensure sufficient jet multiplicity and $b$-tag content for the analysis, we impose a pre-selection that each event must contain:
    \begin{enumerate}
        \item \label{jets} at least six reconstructed jets,
        \item at least four of the jets in \ref{jets} having transverse momenta $p_\text{T} > 40$~GeV, and
        \item at least four of the jets in \ref{jets} being $b$-tagged.
    \end{enumerate}
    Events satisfying these criteria are designated as $4b$ events.  For the purposes of performance evaluation and cross-section limit setting, we further consider the $6b$ samples, defined as events passing the above pre-selection while containing at least six $b$-tagged jets.
    
    For each signal mass point, we prepared one million $4b$ events and ten thousand $6b$ events. For background samples, we prepared one million $4b$ events together with fifty thousand $6b$ events. The $4b$ samples are utilized for training and validation of all neural networks, with the $6b$ datasets reserved for testing and setting the cross-section upper limits.
    
    Since it is much easier to generate a large number of $4b$ events, we train all neural network models on the $4b$ dataset and evaluate them on the $6b$ events. Training directly on $6b$ events, which form a subset of the $4b$ sample, leads to worse performance due to limited event statistics. We demonstrate that models trained on the $4b$ data can generalize well to the $6b$ regime, keeping robust performance for both jet pairing and event classification.
    
    Table~\ref{tab:dataset-summary} summarizes the sizes of the simulated datasets used for each benchmark mass point and the background processes considered in this study. In the $4b$ datasets, 95\% of the events are allocated for neural network training, with the remaining 5\% used for validation. The $6b$ datasets are reserved for evaluating model performance and for studying the cross-section upper limits.
    
    \begin{table}[htbp]
        \centering
        \caption{Summary of the number of simulated events in each category. For the signal samples, the figures are for each of the five benchmark mass points considered here, thus the total numbers in the last row of the table. The $4b$ datasets are used for training and validation, while the $6b$ datasets are reserved for performance evaluation and cross-section upper limit studies.}
        \label{tab:dataset-summary}
        \begin{tabular}{lccc}
            \hline\hline
            Category                & Training $(4b)$ & Validation $(4b)$ & Testing $(6b)$ \\ \hline
            Signal (per mass point) & 950,000         & 50,000            & 10,000         \\
            Background              & 950,000         & 50,000            & 50,000         \\ \hline
            Total                   & 5,700,000       & 300,000           & 100,000        \\
            \hline\hline
        \end{tabular}
    \end{table}

    These samples form the basis for the jet-pairing studies, neural network training, and the statistical limit setting procedure, as described in the following sections.

% section sample_preparation (end)
    
\section{Jet pairing}% (fold)
\label{sec:jet_pairing}

    In the all-hadronic final state of triple Higgs boson production, a large multiplicity of jets arises from the decays of Higgs bosons as well as from initial- and final-state radiation. An important task of the analysis is to correctly identify the pairing of jets originating from the same Higgs boson.
    
    Correct jet pairing enables the reconstruction of Higgs candidates and improves the resolution of relevant observables, such as invariant mass distributions, angular separations, and event shapes. These reconstructed features play a critical role in signal-background classification and have a direct impact on the subsequent analysis.
    
    In this section, we conduct a comparative study of two traditional pairing strategies - the $\chi^2$ method and the absolute mass-difference ($|\chi|$) method - as well as a deep learning-based approach, the \textsc{Spa-Net} method.
    
    \subsection{Traditional pairing methods}% (fold)
    \label{sub:traditional_pairing_methods}
    
        Traditional jet pairing methods define a quantity based on kinematic observables and enumerate all possible combinations of jets to find the configuration that minimizes or maximizes it. These quantities often have clearer physical interpretations, for example, deviations in reconstructed masses from a reference Higgs mass, thereby providing an intuitive understanding of their behavior.
    
        However, such methods require explicit enumeration of all jet pairing configurations. The computational complexity scales as $\mathcal{O}(N_\text{jets}^{n_p})$, where $n_p$ is the number of partons to be matched ($n_p = 6$ in our case). To reduce the number of combinations, the pairing procedure is restricted to the six leading $b$-tagged jets, yielding fifteen distinct possible pairings.
        
        Here, we introduce two conventional pairing methods:
        \begin{itemize}
            \item $\chi^2$ pairing~\cite{Papaefstathiou:2019ofh}: This method minimizes the total squared deviation of the reconstructed Higgs candidate masses from the reference Higgs mass. The $\chi^2$ variable is defined as:
            \begin{equation}
                \chi ^2  = \sum_{i=1}^{3} \left\{ \left[\frac{{m_{h_i}}}{\text{GeV}}\right] - 125 \right\} ^2,
            \end{equation}
            where $m_{h_i}$ is the invariant mass of the $i$-th jet pair. Among fifteen possible pairings, the configuration that yields the lowest $\chi^2$ value is selected.       
        
            \item $|\chi|$ pairing (absolute mass difference): For the method used in the ATLAS analysis~\cite{ATLAS:2024xcs}, the quantity to be minimized is:
            \begin{equation}
                \abs{\left[\frac{m_{h_1}}{\text{GeV}}\right] - 120} + \abs{\left[\frac{m_{h_2}}{\text{GeV}}\right] - 115} +\abs{\left[\frac{m_{h_3}}{\text{GeV}}\right] - 110},
            \end{equation}
            where $m_{h_i}$ are further sorted by their $p_{\text{T}}$'s such that $p_{\text{T}}(h_1) \ge p_{\text{T}}(h_2) \ge p_{\text{T}}(h_3)$. The numbers in this definition are chosen according to the peaks of the $m_{h_i}$ distributions in simulated signal events. The deviation of peaks from the nominated Higgs mass results from the detector effects.
        \end{itemize}
    
    % subsection traditional_pairing_methods (end)
    
    \subsection{\textsc{Spa-Net} pairing}% (fold)
    \label{sub:spanet_pairing}
    
        The Symmetric Preserving Attention Network (\textsc{Spa-Net})~\cite{PhysRevD.105.112008, 10.21468/SciPostPhys.12.5.178, Fenton:2023ikr} is a novel neural network architecture specifically designed for the jet assignment and pairing tasks. By design, \textsc{Spa-Net} can preserve the permutation symmetry in the jet assignment, thereby reducing the number of learnable parameters and eliminating the need for explicit enumeration of all jet permutations.
        
\begin{figure}[t]
            \centering
            {
            \linespread{1.0}
            \begin{tikzpicture}[>=stealth,scale = .93]
            
            \pgfmathsetmacro{\h}{3.0}
            \pgfmathsetmacro{\w}{2.0}

            \coordinate (O) at (0,0);
            \coordinate (Title) at (0, 1.2*\h);
            \coordinate (Input) at (-3.1*\w,0);
            \coordinate (Embedding) at (-1.7*\w, 0);
            \coordinate (CentralTransformer) at (-0.3*\w, 0);
            \coordinate (BranchEncoder) at (1.3*\w, 0);
            \coordinate (TensorAttention) at (2.4*\w, 0);
            \coordinate (Output) at (3.5*\w, 0);
            \coordinate (CTLeftBottomCorner) at ($(CentralTransformer)+(-\w,-\h)$);
            \coordinate (CTRightTopCorner) at ($(CentralTransformer)+( \w, \h)$);
            
            % Central Transformer
            \node[align=center] (CT) at (CentralTransformer |- Title) {Central\\Transformer};
            \draw (CentralTransformer) +(-\w,-\h) rectangle +(\w,\h);
            
            \foreach \x/\n in {-0.75*\w/1, -0.25*\w/2, 0.75*\w/8}{
            \node[draw, rotate=-90, minimum width=1.8*\h cm] at ($(CentralTransformer)+(\x,0)$) {Transformer encoder \n};
            }
            \foreach \i in {-0.2,0,0.2}{
            \fill ($(CentralTransformer)+(0.25*\w,0)$) +(\i,0) circle[radius=2pt];
            }
            
            % Input features
            \node (I) at (Input |- Title) {Inputs};
            \node[draw,minimum width=1.8*\w cm] (I1) at ($(Input)+(0,0.8*\h)$) {$p_{\text{T}1}, \eta_1, \phi_1, m_1, b_1$};
            \node[draw,minimum width=1.8*\w cm] (I2) at ($(Input)+(0,0.2*\h)$) {$p_{\text{T}2}, \eta_2, \phi_2, m_2, b_2$};
            \node[draw,minimum width=1.8*\w cm] (I15) at ($(Input)+(0,-0.8*\h)$) {$p_{\text{T}15}, \eta_{15}, \phi_{15}, m_{15}, b_{15}$};
            
            \node[above] at (I1.north) {Jet Input};
            
            \foreach \i in {0,1,2}{
            \fill (Input) +(0,-0.4*\i - 0.15*\h) circle[radius=2pt];
            }
            
            % Embedding
            \node[align=center] (E) at (Embedding |- Title) {Embedding};
            \foreach \n in {1,2,15}{
            \node[draw] (E\n) at (I\n -| E) {$E$};
            \draw[->] (I\n) -- (E\n);
            \draw[->] (E\n) -- (E\n -| CTLeftBottomCorner);
            }
            
            % Particle Transformers
            \node[align=center] (BE) at (BranchEncoder |- Title) {Particle\\Transformers};
            \node[draw,minimum width=0.65*\w cm,minimum height=0.3*\h cm] (B1) at ($(BranchEncoder)+(0, 0.8*\h)$) {};
            \node[draw,minimum width=0.65*\w cm,minimum height=0.3*\h cm] (B2) at ($(BranchEncoder)+(0,0.25*\h)$) {};
            \node[draw,minimum width=0.65*\w cm,minimum height=0.3*\h cm] (B3) at ($(BranchEncoder)+(0,-0.3*\h)$) {};
            \node (BC) at ($(BranchEncoder)+(0,-0.8*\h)$) {};
            
            \foreach \n in {1,2,3}{
            \node[draw, rotate=-90, minimum width=0.3*\w cm, above=2pt] at (B\n.west) {$T_{1}$};
            \node[draw, rotate=-90, minimum width=0.3*\w cm, below=2pt] at (B\n.east) {$T_{2}$};
            }
            
            % Tensor Attention
            \node[align=center] (TA) at (TensorAttention |- Title) {Outputs};
            \node[draw,minimum width=0.9*\w cm, align=center] (TA1) at (TA|-B1) {Tensor\\attention};
            \node[draw,minimum width=0.9*\w cm, align=center] (TA2) at (TA|-B2) {Tensor\\attention};
            \node[draw,minimum width=0.9*\w cm, align=center] (TA3) at (TA|-B3) {Tensor\\attention};
            \node[draw,minimum width=1.0*\w cm, align=center] (CLS) at (TA|-BC) {Classification\\head};

            \node (O1) at (Output|-TA1) {$h_1 (j_a, j_b)$};
            \node (O2) at (Output|-TA2) {$h_2 (j_c, j_d)$};
            \node (O3) at (Output|-TA3) {$h_3 (j_e, j_f)$};
            \node (OC) at (Output|-CLS) {$\mathcal{S/B}$};
            
            % arrows
            \foreach \i in {1,2,3}{
            \draw[->] (CTRightTopCorner|-B\i) -- (B\i);
            \draw[->] (B\i) -- (TA\i);
            \draw[->] (TA\i) -- (O\i);
            }
            
            \draw[->] (CTRightTopCorner|-CLS) -- (CLS);
            \draw[->] (CLS) -- (OC);
            \end{tikzpicture}
            
            }
            \caption{The high-level model structure of \textsc{Spa-Net}. Each $E$ is an embedding layer, $T_i$ is the transformer encoder, and $h_i$ is the jet assignment result, which contains two jets $j_i$ for the Higgs decay. The particle transformer is a stack of transformer encoders.}
            \label{fig:SPANet_structure}
        \end{figure}
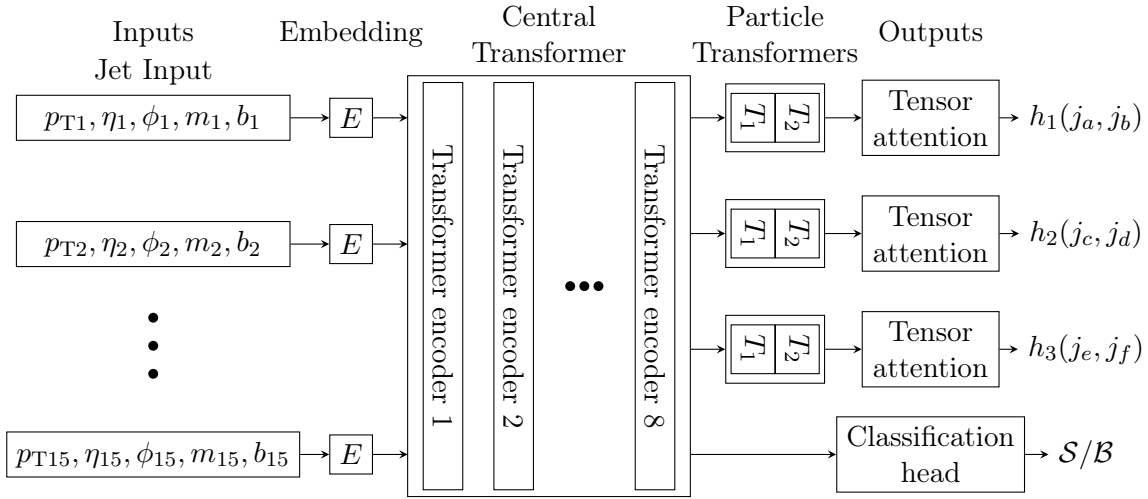

        Figure~\ref{fig:SPANet_structure} presents the high-level model architecture of \textsc{Spa-Net}. Each reconstructed jet is described by its 4-component vector $(p_{\mathrm{T}},\eta,\phi,m)$ together with a boolean $b$-tag, and we keep at most the fifteen leading jets per event. Here, $\eta$, $\phi$, and $m$ correspond to the pseudorapidity, azimuthal angle, and invariant mass of the jet. These input features are encoded in the event-embedding vector by embedding blocks and central transformers. They can encode jet features into an abstract latent space and utilize attention layers to capture the correlations among the jets. The resulting contextual information of jets is then used in subsequent tasks without looping through all possible configurations, in contrast to traditional methods.
        
        In this study, the event embedding is used for both jet assignment and classification tasks. For jet assignment, \textsc{Spa-Net} utilizes the particle transformers and tensor attention mechanisms to further refine the event embedding and construct the jet assignment results. For classification, \textsc{Spa-Net} adopts a basic feed-forward network to distinguish signal events from background events.

        The ground truth jet assignments are defined by matching reconstructed jets to simulated truth quarks within an angular distance of $\Delta R < 0.4$. If multiple jets are matched to the same quark, the closest jet is chosen. If a truth-level quark fails to match any reconstructed jet, the event is still retained in the training samples for jet pairing, unless none of the truth quarks are matched. These matching conditions are applied only during the training of \textsc{Spa-Net}. At inference, \textsc{Spa-Net} is applied to all events passing the pre-selection criteria.
        
        The \textsc{Spa-Net} architecture is designed to preserve key symmetries inherent to the jet assignment task. The event embedding is independent of the jet ordering, due to transformer's properties. Moreover, \textsc{Spa-Net} utilizes the technique of symmetric tensor attention and symmetric loss~\cite{10.21468/SciPostPhys.12.5.178}, ensuring permutation symmetries of labels ({\it e.g.}, the $b \bar b$ and $hh$ pairs) in the output.
        
        In our context, it is essential to emphasize that \textsc{Spa-Net} is not limited to using only the six leading $b$-tagged jets for the jet pairing task, but considers all jets present in an event. This broader scope enables the network to provide correct predictions even in cases where some jets are mistagged or when the Higgs decay jets are not among the six leading $b$-tagged jets. Consequently, \textsc{Spa-Net} is able to leverage a larger dataset for the pairing task compared to traditional methods.
        
        To evaluate the performance of each jet pairing method, we define the event efficiency as the fraction of events in which all three Higgs boson candidates are correctly reconstructed. An individual Higgs candidate is considered correctly reconstructed if the two assigned jets can be matched to the two truth-level $b$-quarks originating from the same Higgs boson.  Only events where all six $b$-quarks are successfully matched to reconstructed jets are included in the event efficiency evaluation. Events with partial matching are excluded from the calculation of event efficiency, although they are retained during training to maximize the usable dataset.

    % subsection spanet_pairing (end)

% section jet_pairing (end)

\section{Neural network classifier}% (fold)
\label{sec:neural_network_classifier}

    We consider two types of neural network classifiers. The first is the dense neural network, representing the basic neural network architecture, which has been employed in the ATLAS tri-Higgs analysis~\cite{ATLAS:2024xcs}, and serves here as our baseline. The second is \textsc{Spa-Net}, which is trained simultaneously on the jet assignment and classification tasks.
    
    \subsection{Dense neural network}% (fold)
    \label{sub:dense_neural_network}
    
        The Dense Neural Network (Dense-NN) is a basic network structure for classification tasks. To distinguish the tri-Higgs signal and background event, we consider the following input features inspired by the ATLAS analysis:
        \begin{itemize} 
            \item $\Delta R_{h_1}, \Delta R_{h_2}, \Delta R_{h_3}$: The angular distance between the two jets form the leading, sub-leading, and least-leading Higgs boson candidates, respectively.
            
            \item $\text{RMS } \Delta R_{\text{dijet}}$: The root mean square of the angular distance between all possible dijet combinations that can form a Higgs boson candidate. We consider all possible permutations of the six jets selected for pairing.
            
            \item $\text{Skewness } \Delta A_{\text{dijet}}$: The skewness of $\cosh\left( \Delta\eta_{ij} \right) - \cos\left( \Delta\phi_{ij} \right) $, where $i,j$ run over all possible dijet combinations that can form a Higgs boson candidate.
            
            \item $H_{\text{T, 6jets}}$: The scalar sum of the $p_{\text{T}}$'s of the six jets that form three Higgs boson candidates.
            
            \item $m_{h}\cos\theta$: $\theta$ is the angle between the reference mass vector $\left( 120, 115, 110 \right) ~\text{GeV}$ and a reconstructed mass vector, defined as $\left( m_{h_1}, m_{h_2}, m_{h_3} \right) - \left( 120, 115, 110 \right) ~\text{GeV}$, where $m_{h_1}, m_{h_2}$, and $m_{h_3}$ are the invariant masses of Higgs boson candidates.
            
            \item $\eta - m_{hhh}$ fraction: It is defined as
            \begin{equation}
                \frac{\sum_{i,j} 2 p_{\text{T},i} p_{\text{T},j} \left( \cosh\left( \Delta\eta_{ij} \right) -1 \right) }{m_{hhh}^2},
            \end{equation}
            where $i,j$ are all possible dijets that can form a Higgs boson candidate and $m_{hhh}$ is the reconstructed tri-Higgs invariant mass.
            
            \item Sphericity and Aplanarity of six jets: We define the momentum tensor $M_{xyz}$ as
            \begin{equation}
                M_{xyz} = \sum_{i}
                \begin{pmatrix}
                p_{xi}^2 & p_{xi}p_{yi} & p_{xi}p_{zi} \\
                p_{xi}p_{yi} & p_{yi}^2 & p_{yi}p_{zi} \\
                p_{xi}p_{zi} & p_{yi}p_{zi} & p_{zi}^2 \\
            \end{pmatrix}
            /
            \sum_{i}\abs{p_i}^2
            \end{equation}
            where $i$ is the jet index. Its eigenvalues are ordered such that $\lambda_1  > \lambda_2 > \lambda_3$. The sphericity is defined as
            \begin{equation}
                S = \frac{3}{2} \left( \lambda_2 + \lambda_3 \right),
            \end{equation}
            and the aplanarity is defined as
            \begin{equation}
                A = \frac{3}{2} \lambda_3.
            \end{equation}
        \end{itemize}
        Features like $H_{\text{T}}$, sphericity, and aplanarity are global event-level variables, which only depend on the selection of the six jets, whereas quantities such as $\Delta R$ additionally rely on how the Higgs boson candidates are reconstructed. Input features are constructed using the different jet pairing methods discussed in section~\ref{sec:jet_pairing} in order to evaluate the robustness of the classification performance with respect to the jet pairing strategy.
        
        We construct three separate $4b$ training datasets using the $\chi^2$, $|\chi|$, and \textsc{Spa-Net} pairing methods. The sample sizes and mass points considered have already been detailed in section~\ref{sec:sample_preparation}. For the \textsc{Spa-Net} pairing, we use the same dataset that was originally employed in training the \textsc{Spa-Net} model itself. 
        
        Although this setup may seem to favor Dense-NN when provided with input features derived from the \textsc{Spa-Net} pairing, we will demonstrate that its classification performance is in fact less sensitive to the pairing strategy. Furthermore, all Dense-NN models continue to underperform significantly compared to the \textsc{Spa-Net} classifier. In principle, constructing  entirely independent datasets for each pairing strategy could further reduce the performance of Dense-NN.
        
        For the $6b$ testing samples, we apply the same pairing strategies used during training, ensuring a consistent evaluation of each classifier under different jet assignment assumptions.
        
        The Dense-NN is implemented using the \verb|Tensorflow| library~\cite{tensorflow2015-whitepaper}. The network architecture consists of fully connected layers with the ReLU activation functions~\cite{agarap2018deep}. Each dense layer is followed by a dropout layer, and the input features are first processed by a batch normalization layer. The model is trained with a binary cross-entropy loss function and optimized using the Adam algorithm~\cite{kingma2015adam}. Early stopping is applied based on the validation loss to mitigate overfitting. In addition, a weighted loss function is employed to address the class imbalance between signal and background samples. The hyperparameters used in the Dense-NN training are summarized in table~\ref{tab:dense_nn_hyperparameter}.
        
        \begin{table}[htbp]
            \centering
            \caption{Hyperparameters used in Dense-NN training.}
            \label{tab:dense_nn_hyperparameter}
            \begin{tabular}{l|l}
            \hline\hline
            Parameter                   & Value \\ \hline
            Number of hidden layers     & 3     \\
            Number of nodes per layer   & 24    \\
            Dropout rate                & 0.1   \\
            Learning rate               & 0.001 \\
            Batch size                  & 2048  \\
            Early stopping patience     & 10    \\
            \hline\hline
            \end{tabular}
        \end{table}
    
    % subsection dense_neural_network (end)
    
    \subsection{\textsc{Spa-Net} Classification}% (fold)
    \label{sub:spanet_classification}
    
        \textsc{Spa-Net} was originally designed for the jet assignment task, but can be naturally extended to perform event-level classification. As described in section~\ref{sub:spanet_pairing}, its architecture produces an event embedding vector that encodes the contextual relationships among jets, which can then be exploited for signal/background classification.
        
        The inputs to \textsc{Spa-Net} consist of low-level quantities: the 4-momentum $(p_{\mathrm{T}}, \eta, \phi, m)$ and $b$-tag flag of each jet. Unlike the high-level physical observables used in Dense-NN, these inputs carry less human-engineered structure, but offer greater flexibility for the network to learn latent representations directly from the data, which can enhance classification performance. As in the case of Dense-NN, \textsc{Spa-Net} is trained on $4b$ events and evaluated on $6b$ events, with the sample sizes and mass points detailed in section~\ref{sec:sample_preparation}.
        
        We adopt a multi-task learning approach in which \textsc{Spa-Net} is trained simultaneously on jet pairing and event classification tasks. Here, we utilize \textsc{Spa-Net} library version 2.3 from Ref.~\cite{spanet_v2.3}. Training is performed with the AdamW optimizer \cite{loshchilov2019decoupled}, including L2 regularization, over 50 epochs. The total loss is defined as  the sum of the jet assignment loss and classification losses, weighted equally. The full hyperparameter settings are summarized in table~\ref{tab:spanet_hyperparameter}.
        
        \begin{table}[htbp]
            \centering
            \caption{Hyperparameters used in \textsc{Spa-Net} training.}
            \label{tab:spanet_hyperparameter}
            \begin{tabular}{l|l}
            \hline \hline
            Parameter                 & Value    \\ \hline
            Learning rate             & 0.0015   \\
            Training epochs           & 50       \\
            Batch size                & 4096     \\
            Dropout                   & 0        \\
            $L_2$ gradient clipping   & 0        \\
            $L_2$ penalty             & 0.0002   \\
            Hidden dimensionality     & 64       \\
            Central encoder layers    & 6        \\
            Branch encoder layers     & 3        \\
            Classification head layers& 2        \\
            Assignment loss weight    & 1        \\
            Classification loss weight& 1        \\
            \hline \hline
            \end{tabular}
        \end{table}
        
        Notably, events with incomplete jet parton matching (i.e., where not all $b$-quarks are matched to jets) are also included in the training process. This inclusion enhances robustness and allows the network to effectively use the training samples.
        
        Although the primary goal at this stage is event classification, incorporating the pairing task during training forces the network to extract more structural information from the event. As a result, the network can build a more expressive embedding. This combined training strategy improves the classification performance compared to training on classification alone.
    
    % subsection spanet_classification (end)

% section neural_network_classifier (end)

\section{Results}% (fold)
\label{sec:results}
    In this section, we present the performance of the jet pairing methods and classification strategies developed for the analysis of triple Higgs boson production. We also report the cross-section upper limits established by various methods.

	\subsection{Jet pairing performance}% (fold)
	\label{sub:jet_pairing_performance}
        
        To evaluate the jet pairing performance, we utilize the event efficiency defined in section~\ref{sec:jet_pairing}. Specifically, we calculate the fraction of events in which all three Higgs bosons are correctly reconstructed. Since \textsc{Spa-Net} is trained on $4b$ samples, we evaluate the pairing performance on both $4b$ and $6b$ datasets to assess its transferability.

        For the traditional pairing methods, we restrict the selection to six jets to control the number of possible configurations. In events with only four or five $b$-tagged jets, we combine them with two or one non-$b$-tagged jet with the highest-$p_{\text{T}}$ to reach six. Conversely, if six or more $b$-tagged jets are present, we select the six with the highest $p_{\text{T}}$. In contrast, \textsc{Spa-Net} uses all available jets in an event and is not limited to $b$-tagged jets. This feature is advantageous, as it avoids the need for special treatment of cases with fewer $b$-tagged jets.

\begin{figure}[htpb]
            \centering
            \subfloat[$4b$ dataset]{  
                \includegraphics[width=0.45\textwidth]{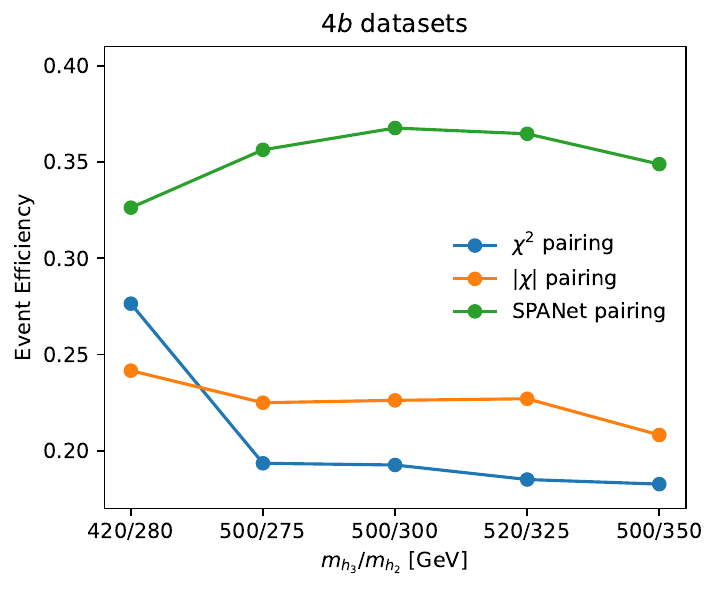}
            }
            \subfloat[$6b$ dataset]{
                \includegraphics[width=0.45\textwidth]{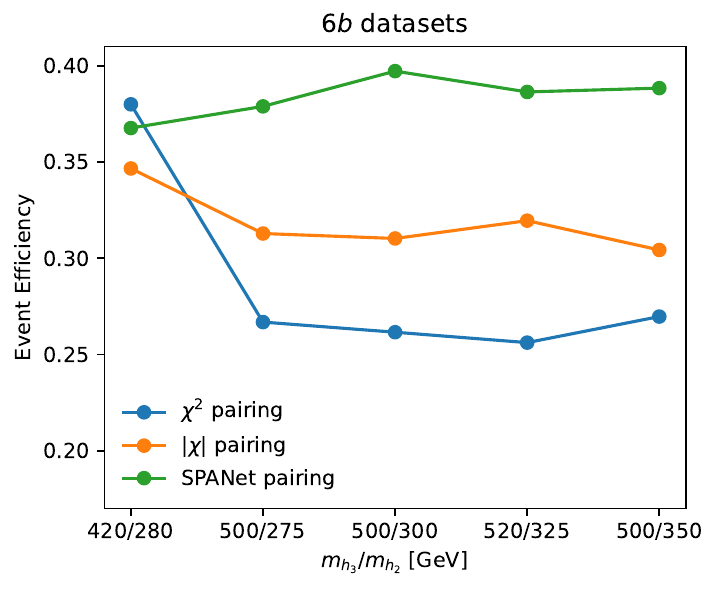}
            }
            \caption{Jet pairing efficiency of various methods across different benchmark mass points.}
            \label{fig:pairing_performance_each_mass_point-TRSM-mix_5-1M}
        \end{figure}

        Figure~\ref{fig:pairing_performance_each_mass_point-TRSM-mix_5-1M} shows the event efficiency for different pairing strategies across the five benchmark mass points. For the $4b$ datasets, \textsc{Spa-Net} consistently achieves the highest efficiency. The $|\chi|$ method yields relatively stable performance across mass points, while the $\chi^2$ method has the lowest efficiency in most cases except at (420, 280)~GeV. For the $6b$ datasets, \textsc{Spa-Net} continues to outperform the traditional methods except at the point (420, 280)~GeV.

        Interestingly, at the mass point (420, 280)~GeV, the $\chi^2$ pairing method slightly outperforms \textsc{Spa-Net} by 1\% in the $6b$ case. This behavior can be understood from the kinematic configuration of this benchmark. In this case, the resonance masses are relatively close, and the intermediate Higgs bosons are not significantly boosted. As a result, the $b$-jets originating from the Higgs decays are more widely separated, leading to more spatially diverse configurations.

        Traditional methods such as $\chi^2$ pairing, which rely solely on invariant mass minimization and remain insensitive to jet angular correlations, are less affected by this type of kinematic diversity. In contrast, \textsc{Spa-Net} may struggle to learn optimal pairing strategies in non-collimated regimes, especially when the available training statistics are limited. Nevertheless, when performance is averaged across all benchmark mass points, \textsc{Spa-Net} consistently achieves higher event efficiencies than traditional methods, underscoring its superior performance in most scenarios.

        We also observe that \textsc{Spa-Net} trained solely on the (420, 280)~GeV benchmark samples achieves better performance than training on the mixed-sample. This suggests that the network has the potential to learn more specialized features and attain higher accuracy when trained on a single benchmark. However, to achieve a similar performance on the mixed-sample case, it would require a substantially larger number of events for training. Given the balance between computational cost and incremental performance improvement, we maintain the mixed-sample training strategy.

        Finally, the $6b$ datasets generally yield better pairing performance than the $4b$ counterparts. This can be explained by the imperfect $b$-tagging information in the $4b$ samples, which introduces greater ambiguity in pairing. Despite being trained on $4b$ events, \textsc{Spa-Net} generalizes well to $6b$ samples, maintaining high pairing accuracy across most mass points. This validates the approach of training on $4b$ datasets, where sample generation is computationally less demanding, while applying the model to $6b$ events in the final analysis. This strategy offers a practical trade-off between performance and computational efficiency.

	% subsection jet_pairing_performance (end)

	\subsection{Signal-background classification performance}% (fold)
	\label{sub:signal_background_classification_performance}
        
        To assess the performance of the signal-background classifiers, we use the Area Under the Receiver Operating Characteristic Curve (AUC) as the evaluation metric. All classifiers are trained on $4b$ datasets and subsequently tested on both $4b$ and $6b$ datasets to examine their generalization capability.

\begin{figure}[htpb]
            \centering
            \subfloat[$4b$ dataset]{  
                \includegraphics[width=0.45\textwidth]{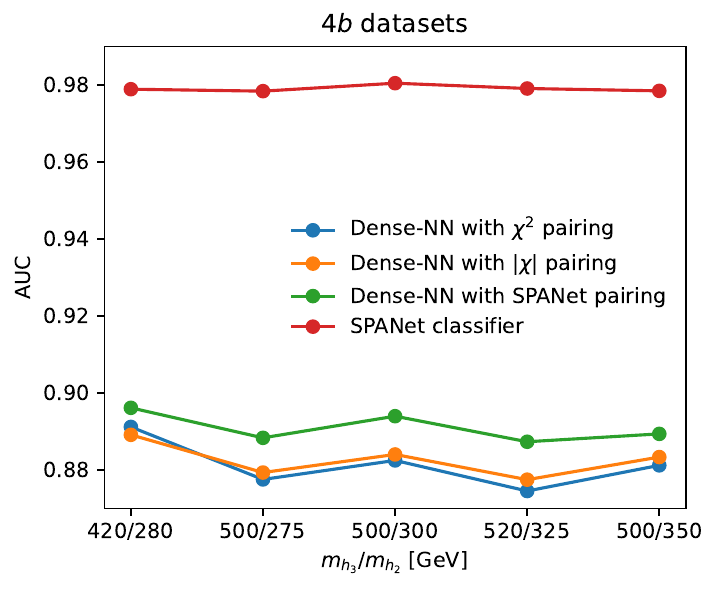}  
            }  
            \subfloat[$6b$ dataset]{  
                \includegraphics[width=0.45\textwidth]{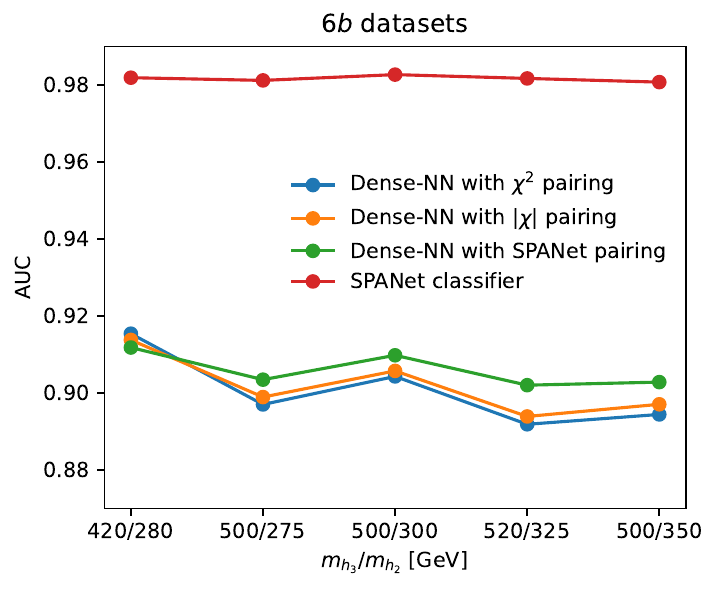}  
            }  
            \caption{Classification performance at different mass points. The AUCs of various classifiers are shown. All models are trained on the combined $4b$ dataset described in section~\ref{sec:sample_preparation}.}
            \label{fig:classification_performance_each_mass_point-TRSM-mix_5-1M}
        \end{figure}

        Figure~\ref{fig:classification_performance_each_mass_point-TRSM-mix_5-1M} shows the classification AUCs across the various benchmark mass points. In all cases, the \textsc{Spa-Net} classifier significantly outperforms the Dense-NN classifiers on both $4b$ and $6b$ datasets. Among the Dense-NN models, the choice of jet pairing method exerts only a minor influence on their overall performance, with all showing broadly similar trends.
        
        The superior performance of \textsc{Spa-Net} stems from both its architectural design and input features. Unlike Dense-NN, which relies on human-engineered high-level observables computed from the six selected jets, \textsc{Spa-Net} operates on low-level jet features for all jets in the event. This richer input allows \textsc{Spa-Net} to learn more expressive representations. In addition, the incorporation of an attention mechanism allows the model to capture complex correlations among jets more effectively.

        We also observe that the Dense-NN models exhibit little sensitivity to the jet pairing strategy.  This result is expected, since many of their input observables are global event-level quantities that do not depend strongly on the specific jet assignments. Consequently, the overall classification performance remains relatively stable across different pairing methods. Nevertheless, the small variations in AUC across pairing strategies track the corresponding jet pairing efficiencies, suggesting that higher-quality pairing can still offer incremental training benefits, albeit with a weak dependence.
        
        When comparing performance between $4b$ and $6b$ evaluations, we observe that the $6b$ datasets consistently yield better results across all mass points than the $4b$ datasets. This behavior mirrors the improved pairing performance discussed in section~\ref{sub:jet_pairing_performance} and further supports the validity of training neural networks on $4b$ events while applying them to $6b$ samples in the final analysis.

	% subsection signal_background_classification_performance (end)

	\subsection{Cross-section upper limits}% (fold)
	\label{sub:cross_section_upper_limits}

        In this section, we present the cross-section upper limits on triple Higgs boson production for the benchmark mass points under study. The statistical analysis follows the $\text{CL}_{\text{s}}$ method~\cite{Read:2002hq}, with the test statistic defined as the output score from the neural network classifier.

        The classifier scores are evaluated on the $6b$ testing datasets and binned uniformly into 20 intervals over the range $[0,1]$. Event yields are scaled to  reference luminosities of 300~fb$^{-1}$ or 3000~fb$^{-1}$. Systematic uncertainties are not included in this study.

        The parameter of interest (POI) is the signal strength $\mu_s$, defined as the ratio of the fitted signal cross-section to the true signal value. The POI is excluded at the 95\% confidence level if $\text{CL}_{\text{s}} < 0.05$. We use the \verb|pyhf|~\cite{pyhf,pyhf_joss} package to calculate the upper limits. Once an signal strength upper limits are obtained, they are translated into the corresponding cross-section upper limits.  
        
        Table~\ref{tab:cutflow} summarizes the pre-selection results for two representative benchmark mass points. Here, the Efficiency refers to the fraction of events passing a given selection step relative to those passing the previous step, while the Passing Rate reflects the total fraction of events that survive all cuts up to that stage. Other benchmark points exhibit cutflow patterns similar to the (500, 300)~GeV case and are therefore omitted. Combined with table~\ref{tab:trsm_xsec}, we can obtain the absolute cross-sections after pre-selection for use in the upper limit computation.

        \begin{table}[htpb]
            \centering
            \caption{Cutflow table for the $(m_{h_3}, m_{h_2}) = $ (420, 280)~GeV and (500, 300)~GeV mass points. The cut efficiency denotes the fraction of events passing a given cut among those that pass the previous one, while the passing rate represents the fraction of events that pass all cuts up to that stage relative to the initial number of events.}
            \label{tab:cutflow}
            \begin{tabular}{c|cc|cc}
                \hline \hline
                Cuts & \multicolumn{2}{c|}{(420, 280)~GeV} & \multicolumn{2}{c}{(500, 300)~GeV} \\
                     & Efficiency & Passing Rate & Efficiency & Passing Rate \\ \hline
                $n_{\text{jet}} \ge 6$           & 0.608 & 0.608 & 0.684 & 0.684 \\
                $n_{\text{jet}}(p_{\text{T}} > 40\,\text{GeV}) \ge 4$      & 0.820 & 0.499 & 0.891 & 0.610 \\
                $n_{b\text{-jet}} \ge 4$         & 0.642 & 0.320 & 0.671 & 0.409 \\
                \hline \hline
            \end{tabular}
        \end{table}

        All signal and background samples are generated at $\sqrt{s}=13$~TeV. To obtain projections at 14~TeV, we rescale the expected event yields using the ratio of LO cross-sections listed in table~\ref{tab:trsm_xsec}. Since the kinematic distributions of Higgs production at 13 and 14~TeV are very similar, this procedure provides a reliable extrapolation to HL-LHC conditions at 14~TeV with integrated luminosities of $\mathcal{L}=300$ and $3000~\text{fb}^{-1}$.

\begin{figure}[htpb]
            \centering
            \subfloat[14~TeV, $\mathcal{L} = 300~\text{fb}^{-1}$]{  
                \includegraphics[width=0.45\textwidth]{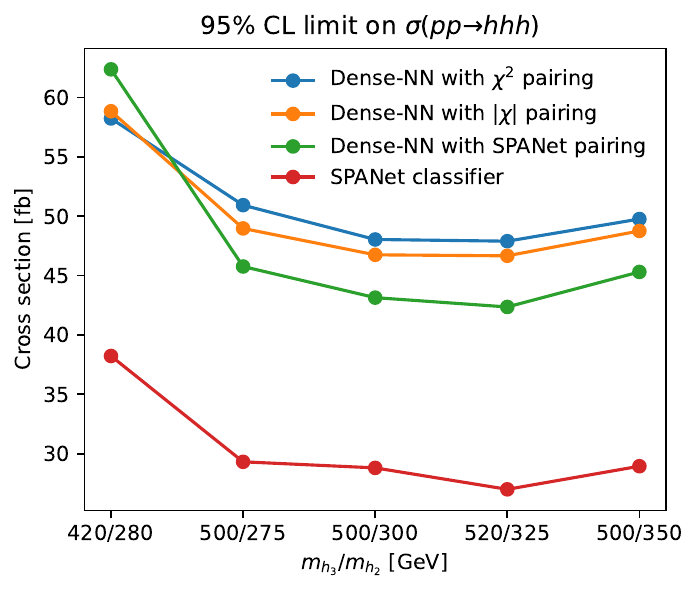}  
            }
            \subfloat[14~TeV, $\mathcal{L} = 3000~\text{fb}^{-1}$]{  
                \includegraphics[width=0.45\textwidth]{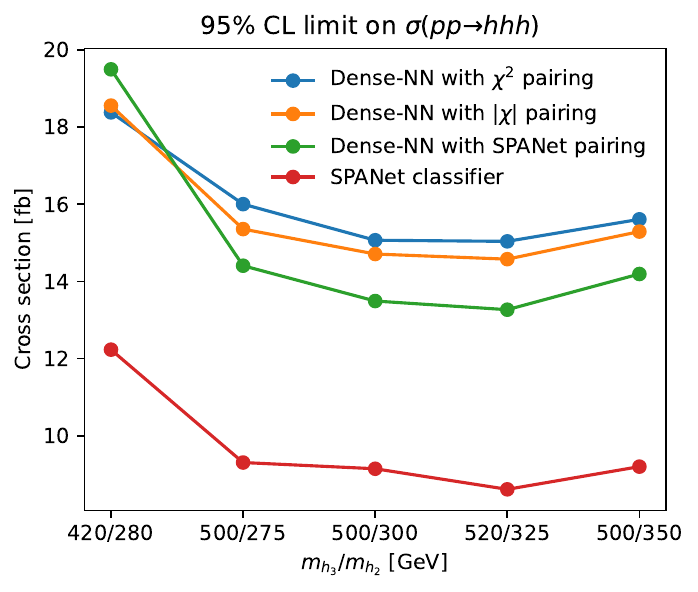}  
            }
            \caption{Expected 95\%~CL upper limits on the triple Higgs production cross-section, $\sigma(pp \to hhh)$, at the 14-TeV LHC for different classifiers and luminosities. The limits are computed using the $\text{CL}_{\text{s}}$ method.}
            \label{fig:cross_section_upper_limit-TRSM-mix_5-1M}
        \end{figure}

        Figure~\ref{fig:cross_section_upper_limit-TRSM-mix_5-1M} shows the expected 95\%~CL upper limits on $\sigma(pp\to hhh)$ for various benchmark points in the TRSM. We present the results under different collider configurations. Dense-NN classifiers using different jet pairings yield similar results, consistent with their comparable classification performance. In contrast, the \textsc{Spa-Net} classifier provides the strongest exclusion power across all benchmark mass points, improving the expected upper limits by up to 40\% relative to the Dense-NN classifiers.
        
        In these high-luminosity scenarios, the projected upper limits approach the expected signal cross-sections, highlighting the potential to observe or place stringent constraints on resonant tri-Higgs production at the HL-LHC with the aid of advanced machine learning techniques such as \textsc{Spa-Net}.

        These results demonstrate both the effectiveness of the \textsc{Spa-Net} architecture and the advantages of the combined training strategy. The approach significantly enhances the sensitivity to the triple-Higgs signal and offers a promising framework for future searches in this challenging but critical channel.

	% subsection cross_section_upper_limits (end)

% section results (end)

\section{Conclusions}% (fold)
\label{sec:conclusions}

    In the analysis of tri-Higgs production in the the all-hadronic $6b$ final state, the workflow centers on two critical tasks: jet pairing and event classification. Due to the highly complex final state and the limitations of imperfect $b$-tagging performance, traditional jet-pairing approaches exhibit reduced efficiency. In the classification part, Dense-NNs approaches require the construction of high-level input features accurately, and their performance remains sensitive to the specific jet pairing method employed.

    In this work, we demonstrate that the \textsc{Spa-Net} architecture provides a significantly more effective solution. On the one hand, \textsc{Spa-Net} is specifically designed to address the combinatorial complexity of the jet assignment problem. Its architecture preserves the permutation symmetries of jet and Higgs label exchanges and can naturally handle cases with mis-tagged jets. On the other hand, it enables the combination of the training of jet pairing and event classification, allowing the network to build more suitable representations that benefit both tasks simultaneously.

    For the jet pairing task, we trained a \textsc{Spa-Net} model on mixed-mass $4b$ datasets. We found that its event pairing efficiency surpasses those of the classical $\chi^2$ and $|\chi|$ methods across most benchmark mass points. When applied to $6b$ test samples, \textsc{Spa-Net} pairing still outperforms the traditional methods except at the $(420,280)$~GeV point. Notably, we further observed that its pairing performance improves with larger training sample sizes, suggesting that increased statistics can mitigate challenges associated with ambiguous kinematic configurations..

    For event classification, we trained both Dense-NN and \textsc{Spa-Net} classifiers using the $4b$ datasets and evaluated them on $6b$ events. The \textsc{Spa-Net} classifier consistently outperformed the Dense-NNs. We attribute this improvement to two main factors: the shared embedding architecture, which benefits from simultaneous training on jet assignment and classification, and the use of low-level jet features, which allow \textsc{Spa-Net} to learn more expressive event representation.

    Finally, we used the neural network output scores as test statistics for the $\text{CL}_{\text{s}}$ method to estimate expected upper limits on the triple-Higgs production cross-section. The \textsc{Spa-Net} classifier yields significantly stronger constraints, improving the expected sensitivity by up to 40\% compared to Dense-NNs.

    In conclusion, this study highlights the potential of \textsc{Spa-Net} as a powerful tool for triple-Higgs searches in the $6b$ channel. By integrating jet assignment and classification within a single architecture, \textsc{Spa-Net} enhances both reconstruction and discrimination, leading to significantly tighter cross-section limits. These results demonstrate the strong potential of advanced machine learning methods to expand discovery reach and improve sensitivity for multi-Higgs searches at the HL-LHC.

% section conclusions (end)

\section*{Acknowledgments}
This work was performed in part at the Aspen Center for Physics, which is supported by a grant from the Simons Foundation (1161654, Troyer).  CWC and FYH were supported by the National Science and Technology Council under Grant Nos. NSTC-111-2112-M-002-018-MY3 and NSTC-114-2112-M-002-020-MY3. IL is supported in part by the U.S. Department of Energy, Office of High Energy Physics, under contracts DE-AC02-06CH11357 at Argonne and DE-SC0010143 at Northwestern. SCH are supported by the U.S. National Science Foundation under Grant Number 2209034. SC and IL also acknowledge the hospitality of the National Center for Theoretical Sciences at National Taiwan University, where part of this work was completed.

\appendix

\bibliographystyle{utphys}
\bibliography{reference}

\end{document}